\documentclass[twocolumn,showpacs,preprintnumbers,amsmath,amssymb,superscriptaddress]{revtex4-1}

\usepackage{graphicx}
\usepackage{dcolumn}
\usepackage{bm}
\usepackage{tabularx}
\usepackage{makecell}
\usepackage{braket}
\usepackage{lipsum}
\usepackage{ulem}
\usepackage{amsmath,amssymb} 
\newcolumntype{M}{>{\centering\arraybackslash}m{1.85cm}}
\usepackage[export]{adjustbox}
\usepackage{longtable}
\usepackage{float}
\usepackage{xcolor}
\usepackage{color}   
\usepackage{hyperref}
\hypersetup{
	colorlinks=true, 
	linkcolor=blue,  
}

\makeatletter
\newcommand{\colorcaption}[2][]{%
	\begingroup%
	\renewcommand{\@caption@fignum@sep}{ (Color online). }%
	\caption[#1]{#2}%
	\endgroup%
}
\makeatother

\begin{document}

\title{Higher forbidden unique $\beta^-$ decay transitions and shell-model interpretation}

\author{Archana Saxena}
\affiliation{Department of Physics, Indian Institute of Technology Roorkee, Roorkee 247667, India}
\author{Praveen C. Srivastava}
 \email{praveen.srivastava@ph.iitr.ac.in}
\affiliation{Department of Physics, Indian Institute of Technology Roorkee, Roorkee 247667, India}

\begin{abstract}
In the present work, we have predicted the half-lives for the $\beta^{-}$ decay for higher forbidden unique transitions in the mass range of nuclei from A = 40-138. For these transitions, the experimental data for half-lives are not available except for a few cases. The calculations for half-lives are performed within the framework of the nuclear shell model  (SM). We have used the effective interactions sdpf-mu, gxpf1a, gwbxg, G-matrix, snet, sn100pn, and jj56pnb to perform the SM calculations in different mass regions. A comprehensive discussion has been made between the SM predicted half-lives and the scaled half-lives from proton-neutron quasiparticle random-phase approximation (pnQRPA).
 The results of the present study will be  useful to plan new experiments to measure half-lives for these higher forbidden unique $\beta^{-}$ transitions.
\end{abstract}

\maketitle

{\color{black}
	\section{Introduction}
The nuclear $\beta$-decay, also known as the weak nuclear processes occurring inside the nuclei, is a powerful tool for exploring the structure and properties of atomic nuclei. It plays an important role in subatomic physics and  astrophysics \cite{review1,review01,review11,review111,review2,review3,review4,Anil,Shweta1,Shweta2}.
To study the $\beta$-decay, we need a theoretical model that can explain the properties of the nuclei. In the present work, we  use the nuclear shell model {\color{black}(SM)} to calculate the matrix elements involved in beta decay for {\color{black}higher} forbidden unique transitions. The decay rates depend on the precisely calculated wave functions of parent and daughter nuclei involved in nuclear matrix elements (NMEs) \cite{Suhonen1}.
 
The transitions in the $\beta$-decay can be categorized into two types, allowed and forbidden, based on the orbital angular momentum ($l$) of the emitted leptons.
 When  $l=0$, it is categorized as  allowed, while  $l > 0$ leads to forbidden transitions.
 Allowed transitions are characterized by a change in total angular momentum $\Delta J$ = 0, $\pm$1 and no change in parity ($\Delta\pi$ = +1). The forbidden transitions are further categorized as forbidden non-unique (FNU) and forbidden unique (FU). {\color{black} For the $K$th-FNU transition, $\Delta J = K$ (In case of K=1, $\Delta J = 0,1$), where $K$ is the order of forbiddenness} and parity changes in the odd-forbidden and remains same in the even-forbidden decays according to $\Delta\pi$ = $(-1)^K$. Basically, forbidden unique decays are a special case of forbidden non unique (FNU) decays. In the $K$th FU transition, $\Delta J$ = $K+1$, here, the angular momentum change between the initial and final nuclear states is maximal and parity follows the same as in FNU decays. Consequently, the decay rate and the shape factor are significantly simplified as the contribution of only one matrix element \cite{jouni}. 

 The majority of $\beta$-decay  transitions observed in nature are of the allowed type and a few with a low order of `$K$'. However, recent research has focused on finding rare beta decays that have longer {\color{black}half-lives}. These transitions are  hindered {\color{black}by} low Q value or high order of {\color{black}`$K$'}. 
The theoretical formalism of nuclear beta decay is explained in the book by Behrens and B$\ddot{{\rm u}}$hring \cite{behrens1982}.
 With advancements in measurement technology, the study of {\color{black}higher} forbidden beta decays became interesting.
Recently, half-lives for 148 forbidden unique beta transitions, including 2nd, 3rd, 4th, 5th, 6th, and 7th orders, have been predicted using the {\color{black}proton-neutron quasiparticle random-phase approximation (pnQRPA)} model by Kostensalo and Suhonen \cite{joel12017}. The experimental half-lives for these {\color{black}higher} forbidden unique transitions are not  {\color{black}yet} available.

The forbidden unique decays exhibit a straightforward dependence on the weak axial-vector coupling constant $g_{A}$ ($t_{1/2} \propto g_{A}^{-2}$). For bare nucleons, the value  $g_{A} = 1.27 $ is extracted from the partially conserved axial-vector current hypothesis (PCAC) but this value is affected inside the nuclear matter due to nuclear medium eﬀects or nuclear many-body effects and a quenching is needed to reproduce the experimental data. So, the role of  $g_{A}$ is important to  calculate half-lives  more precisely. There are few methods available to study the quenching in $g_{A}$ values in higher forbidden unique $\beta$- decay. Previously, the effective value of $g_{A}$ has been investigated through a half-life comparison method, where predicted and experimental values are compared across various $g_{A}$ values \cite{Ejiri2014, Ejiri2015,joel22017}. This approach can be used to find out the quenching once experimental data becomes accessible. The spectrum-shape method (SSM) is the popular one to study the quenching in which the shapes of calculated and experimental electron spectra are compared \cite{ joel22017, anil2021, mika2016}. In the present manuscript, we have done a systematic SM study for the {\color{black}higher} forbidden unique  $\beta^{-}$ transitions to test the predictive power of our computed nuclear wave functions in the mass region A = 40-138.

The present manuscript is structured into the following sections. The section \ref{formalism} is divided into two subsections: {\color{black} subsection \ref{beta} explains the theoretical formalism used to calculate the shape factor, nuclear matrix elements (NMEs), and half-lives, while subsection \ref{Hamiltonian} reported the shell model Hamiltonian employed}. Section \ref{result} presents the SM results for half-lives of higher FU  $\beta^{-}$ decay transitions. Finally, we conclude the manuscript in Section \ref{Conclusion}.

\section{Theoretical Framework} \label{formalism}

\subsection{ Formalism} \label{beta} 
In the present work, we have calculated the partial {\color{black}half-lives  for the forbidden unique $\beta^{-}$ transitions}. The generalized  theoretical formalism of nuclear $\beta$-decay is available in Refs \cite{mika2017, mst2006}. {\color{black}For allowed and forbidden types of $\beta$-decay, the detailed formalism is given in the books by  Behrens and B$\ddot{{\rm u}}$hring \cite{behrens1982} and by  Schopper \cite{hfs1966}.}
The formalism of nuclear $\beta$-decay  theory is based on the impulse approximation which says that the decaying nucleon doesn't interact with the other nucleons around it at the moment of decay. However, before and after the decay, it interacts with these nucleons strongly. This simplified approach helps in understanding and calculating both allowed and forbidden {$\beta$-decay processes.


   The expression for calculating the partial half-life is given as \cite{mika2017, behrens1982}

    \begin{equation}
	t_{1/2}= \frac{\kappa}{\tilde{C}},
        \end{equation}
 
	where $\kappa$ is a constant and value is given as \cite{Patrignani}

\begin{eqnarray}
\kappa =
\frac{2\pi ^{3}\hbar ^{7}{\mathrm{ln(2)}}}{m_{e}^{5}c^{4}(G_{\mathrm{F}}{\mathrm{cos}}\theta _{\mathrm{C}})^{2}}=6289~s,
\end{eqnarray}
	where, $G_{F}$ is the Fermi constant and $\theta_{\mathrm{C}}$ is the Cabibbo angle. The $\tilde{C}$ is the dimensionless integrated shape function and can be calculated as
  
  \begin{eqnarray}
\label{tc}
\tilde{C}=\int _{1}^{w_{0}}C(w_{e})pw_{e}(w_{0}-w_{e})^{2}F_{0}(Z,w_{e})dw_{e}.
\end{eqnarray}
In the above expression the  usual dimensionless kinematics quantities divided by the electron rest mass are used as $w_0=W_0/m_ec^2$, $w_e=W_e/m_ec^2$, and $p=p_ec/m_ec^2=\sqrt{(w_e^2-1)}$. The $w_0$ is the endpoint energy,  $w_e$  and $p_e$ are the total energy and momentum of emitted electrons, respectively. The term  $F_0(Z, w_e)$ is the Fermi function, and  $Z$ is the atomic number of the daughter nucleus. The  $C(w_e)$ is the shape factor that contains information about the nuclear structure.  

 The general form of shape factor $C(w_{e})$ is given as:
 \begin{eqnarray} \label{eq2}
C(w_e)  = \sum_{k_e,k_\nu,K}\lambda_{k_e} \Big[M_K(k_e,k_\nu)^2+m_K(k_e,k_\nu)^2 \nonumber\\
    -\frac{2\gamma_{k_e}}{k_ew_e}M_K(k_e,k_\nu)m_K(k_e,k_\nu)\Big],
\end{eqnarray}
 
where, $k_e$ and $k_\nu$ (= 1, 2, 3,...) are positive integers and associated with the partial-wave expansion of the leptonic wave functions, and $K$ denotes the  order of forbiddenness of the $\beta$ decay. The terms  $M_K(k_e,k_\nu)$ and $m_K(k_e,k_\nu)$ can be expressed in  terms of different {\color{black}NMEs} and other kinematic factors \cite{mika2017, behrens1982}. 
Here, $	\gamma_{k_e}  =\sqrt{k_e^2-(\alpha{Z})^2}$ and  $y=(\alpha{Zw_e}/p_ec)$ are the auxiliary quantities, and  $\alpha=1/137$ is the fine structure constant and 
the term $\lambda_{k_e}={F_{k_e-1}(Z,w_e)}/{F_0(Z,w_e)}$,
is the  Coulomb function, in which ${F_{k_e-1}(Z,w_e)}$ is the generalized Fermi function \cite{mika2017, mst2006, Anil2020PRC} which can be calculated using formula
\begin{eqnarray} \label{eq3}
F_{k_e-1}(Z,w_e) &=4^{k_e-1}(2k_e)(k_e+\gamma_{k_e})[(2k_e-1)!!]^2e^{\pi{y}} \nonumber \\
 & \times\left(\frac{2p_eR}{\hbar}\right)^{2(\gamma_{k_e}-k_e)}\left(\frac{|\Gamma(\gamma_{k_e}+iy)|}{\Gamma(1+2\gamma_{k_e})}\right)^2.
\end{eqnarray}

NMEs contain all the information about nuclear-structure and can be expressed as

\begin{align}
\begin{split}
^{\rm V/A}\mathcal{M}_{KLS}^{(N)}(pn)(k_e,m,n,\rho)& \\ =\frac{\sqrt{4\pi}}{\widehat{J}_i}
\sum_{pn} \, ^{\rm V/A}m_{KLS}^{(N)}(pn)(&k_e,m,n,\rho)(\Psi_f|| [c_p^{\dagger}
\tilde{c}_n]_K || \Psi_i).
\label{eq:ME}
\end{split}
\end{align}
Here, summation runs over the proton (p) and neutron (n) single particle states.
The functions ${^{V/A}m_{KLS}^{(N)}}(pn)(k_e,m,n,\rho)$ are the  single particle matrix elements (SPMEs){, \color{black}  they do not depend on the nuclear models. The terms $(\Psi_f|| [c_p^{\dagger}\tilde{c}_n]_K || \Psi_i)$ are the one-body transition densities (OBTDs) between the initial $(\Psi_i)$ and final $(\Psi_f)$ nuclear states and these are model-dependent}. 
In the present work, the SPMEs  use harmonic oscillator wave functions  \cite{mika2017,mst2006}. The OBTDs are calculated using nuclear shell model codes KSHELL \cite{KSHELL} and  NUSHELLX \cite{nushellx}.
We have also included here the next-to-leading-order (NLO) corrections to the shape factor which increase the number of NMEs (full details about the NLO corrections are given in Refs. \cite{mika2016,mika2017}).

 \subsection{Shell-Model Hamiltonian} \label{Hamiltonian}
 We have performed  calculations within the framework of shell-model. The  nuclear shell-model Hamiltonian can be written in terms of single particle energies and two nucleon interactions as
\begin{equation}
H = T + V = \sum_{\alpha}{\epsilon}_{\alpha} c^{\dagger}_{\alpha} c_{\alpha} + \frac{1}{4} \sum_{\alpha\beta \gamma \delta}v_{\alpha \beta \gamma \delta} c^{\dagger}_{\alpha} c^{\dagger}_{\beta} c_{\delta} c_{\gamma},
\end{equation}
where $\alpha = \{n,l,j,t\}$ represents a single-particle state  and   $\epsilon_{\alpha}$ is the single particle energy of state $\alpha$ and   $c^{\dagger}_{\alpha}$ and $c_{\alpha}$ are the fermion pair creation and annihilation operators. The term
$v_{\alpha \beta \gamma \delta} = \langle\alpha \beta | V | \gamma \delta\rangle $ represents the antisymmetrized two-body matrix element (TBME).

We have taken different effective interactions for different model spaces for nuclei ranging from mass  $A = 40 - 138$. 
The effective interactions sdpf-mu \cite {SDPFMU}, gxpf1a\cite{gxpf1a}, gwbxg \cite{gwbxg1,gwbxg2}, G-matrix \cite{G-matrix}, snet \cite{snet1,snet2}, sn100pn \cite{sn100pn1,sn100pn2} and jj56pnb \cite{Entem,jj56pnb}  are used for the calculations and  shown in last column of Tables \ref{table1}-\ref{table4}.

In the case of sdpf-mu effective interaction we have used full sdpf model space with $^{16}$O as a core. The valence model space has the orbitals  $1d_{5/2}$, $2s_{1/2}$, $1d_{3/2}$ $1f_{7/2}$, $2p_{3/2}$, $1f_{5/2}$ and $2p_{1/2}$. For the generation of a negative parity state in $^{40}$K, we have used a truncation of 1p-1h excitation from $sd$ to $pf$ shell. Here, we are exciting only one particle either proton or neutron, while 2p-2h calculations have been done for $^{40}$Ca case.

We have used gxpf1a effective interaction for nuclei having mass range from  $A=40-62$ and the calculations have been done  without truncation in full $pf$ model space and $^{40}$Ca is considered as a core. Here, the orbitals for valence protons and neutrons are 1$f_{7/2}$, 2$p_{3/2}$, 1$f_{5/2}$, 2$p_{1/2}$ with single particle energies (SPEs) (in MeV) are $-8.6240$, $-5.6793$, $-1.3829$, $-4.1370$.

The model space of gwbxg effective interaction consists of
1$f_{5/2}$, 2$p_{3/2}$, 2$p_{1/2}$, 1$g_{9/2}$  proton orbitals and 
2$p_{1/2}$, 1$g_{9/2}$, 1$g_{7/2}$, 2$d_{5/2}$, 2$d_{3/2}$, and 3$s_{1/2}$  neutron orbitals 
and  SPEs (in MeV) used in this interaction are 
$-5.322 , -6.144, -3.941, -1.250$, for the proton orbitals, and $ -0.696,  -2.597,
+5.159, +1.830, +4.261, +1.741 $ for the neutron orbitals, respectively using  $^{68}$Ni  as a core. This effective interaction is prepared with different interactions and the original 974 two-body matrix elements (TBMEs) are obtained from the bare G-matrix of the H7B potential [H7B] \cite{H7B}, for more details, see Refs \cite{Ji1988, Gloeckner1975, Serduke1976}.
 For $A=84-92$ we apply 1p-1h truncation across $N=50$ shell, and from mass A=94 to 104 we completely filled 1g$_{9/2}$ orbital and open all four neutron orbitals above $N=50$ .

Corresponding to G-matrix interaction, the model space of the present study consists of two proton orbitals 2$p_{1/2}$ and 1$g_{9/2}$ with {\color{black}SPEs} (in MeV) are $0.000, +0.9000$ and five neutron orbitals 2$d_{5/2}$, 3$s_{1/2}$, 2$d_{3/2}$, 1$g_{7/2}$, and 1$h_{11/2}$   with SPEs are $0.000, +1.2600, +2.6300, +2.230, +3.500$ respectively. Here, $^{88}$Sr is taken as inert core and we have not employed any truncation in the present calculations. The realistic eﬀective interaction obtained by the G-matrix approach, which was constructed by the eﬀective microscopic interaction derived from a charge-symmetry breaking nucleon-nucleon potential \cite{Machleidt2001} with further modiﬁcations in the monopole part and was used in Ref. \cite{G-matrix}.

The snet effective interaction consists of 17 orbitals, eight are proton orbitals and nine are neutron orbitals. 

In the calculations using snet effective interaction, we have put a minimum of 8-10 protons in 1$g_{9/2}$ orbital and a maximum of 2 protons in 1$g_{7/2}$  orbital, while neutrons are occupying the 1$g_{7/2}$, 2$d_{5/2}$, 2$d_{3/2}$, 3$s_{1/2}$ and 1$h_{11/2}$ orbitals. Further, for computational reasons, we have restricted a minimum of 2-6  neutrons in the 1$h_{11/2}$  orbital from A=114 to 124, respectively.

The shell-model eﬀective interaction sn100pn is made up of  1$g_{7/2}$, 2$d_{5/2}$, 2$d_{3/2}$,  3$s_{1/2}$, 1$h_{11/2}$, orbitals for protons and neutrons in the $50-82$ model space. The  SPEs (in MeV) for neutrons orbitals are $-10.609$, $-10.289$, $-8.717$, $-8.694$, and $-8.815$  corresponding to  the 1$g_{7/2}$, 2$d_{5/2}$, 2$d_{3/2}$, 3$s_{1/2}$, and 1$h_{11/2}$  neutron orbitals, respectively. The  SPEs (in MeV) for protons orbitals are $+0.8072$, $+1.5623$, $+3.3160$, $+3.2238$, and $+3.6051$  corresponding to neutron orbitals  1$g_{7/2}$, 2$d_{5/2}$, 2$d_{3/2}$,  3$s_{1/2}$, 1$h_{11/2}$, respectively. The residual two-body interaction starts with a G-matrix derived from the CD-Bonn nucleon-nucleon interaction \cite{sn100pn2}. This sn100pn interaction is designed from Ref. \cite{sn100pn1}. We have not applied any truncation where sn100pn interaction is used.

The jj56pnb \cite{jj56pnb} effective interaction {\color{black}is} created for valence space where proton numbers $(Z)$ ranging from 50 to 82 and neutron numbers $(N)$ ranging from 82 to 126. This interaction is derived from N3LO interaction \cite{Entem} and  the Coulomb interaction part is included in the proton-proton interaction part. The proton orbitals $1g_{7/2}$ , $2d_{5/2}$ , $2d_{3/2}$, $3s_{1/2}$ , $1h_{11/2}$  {\color{black}have SPEs (in MeV)}  $-9.667$, $-8.705$, $-6.959$, $-7.327$, $-6.874$ and neutron orbitals  {\color{black}}  $1h_{9/2}$ , $2f_{7/2}$ , $2f_{5/2}$ , $3p_{3/2}$ , $3p_{1/2}$ , $1i_{13/2}$  have $-0.842$, $-2.4030$, $-0.3980$, $-1.549$, $-1.040$, $+0.2970$, respectively. Using  this interaction we have not used any truncation in our calculations.

\begin{table*}

		\centering

		\caption{\color{black}Comparison of shell model half-lives for 2nd-, 3rd-, 4th-, 5th-, 6th- and 7th- FU $\beta^{-}$ transitions  with coupling constants $g^{eff}_{A}$=1.0, $g_{V}$=1.0, and half-lives using pnQRPA approach with $g^{eff}_{A}$=1.0 \cite{joel12017}. The ground-state to ground-state Q value is taken from AME2020 \cite{AME2020}. Here, the $0^{+}$ is the ground state of an even-even nucleus and the asterisk (*) represents the involved isomeric states.\label{table1}}
		
  \begin{ruledtabular}
			\begin{tabular}{lcccccc}
				 ~~~~~ Transition & Forbiddenness & {\color{black}Q value} & & \multicolumn{2}{c}{Partial half-life}  & Interaction \\
				  
				  \hline

			~~~~~ $\beta^-$ Decay & Kth FU &(keV)  &  & {\color{black}SM} & {\color{black}Ref \cite{joel12017}} \\
			
			\hline

 $^{50}$Ca$(0^+) \rightarrow $$^{50}$Sc($5^+$) & 4th FU  & $4947.9(30)$  &     &$2.8 \times 10^{5}$ yr& $3.3(5) \times 10^{5}$ yr & gxpf1a \\

  $^{50}$Sc$(5^+) \rightarrow $$^{50}$Ti($0^+$) & 4th FU  & $6894.7 (25)$ &   &$1.0\times 10^{5}$ yr &  $1.3(2) \times 10^{5}$ yr	& gxpf1a \\

    $^{52}$V$(3^+) \rightarrow $$^{52}$Cr($0^+$) & 2nd FU  & $3976.48 (16)$ &    & $1.2$ yr	 & $7.3(9) \times 10^{3}$  yr  &gxpf1a \\
    
    $^{52}$Ti$(0^+) \rightarrow $$^{52}$V($3^+$) & 2nd FU  & $1965.3(28)$ & & $ 4.0 \times 10^{1}$ yr & $ 1.1(20) \times 10^{2}$ yr &gxpf1a \\
    
   $^{52}$Sc$(3^+) \rightarrow $$^{52}$Ti($0^+$) & 2nd FU  & 8954 (4)  & & 54.3 h & 34.4(5) h & gxpf1a \\

$^{54}$V$(3^+) \rightarrow $$^{54}$Cr($0^+$) & 2nd FU  & 7037 (11)  &   &    24.6 d & 9.4(2) d 
& gxpf1a \\

$^{54}$Mn$(3^+) \rightarrow $$^{54}$Fe($0^+$) & 2nd FU  & 696.4(11) &     &$6.5 \times 10^{5}$ yr  &$6.3(9) \times 10^{5}$ yr  &gxpf1a \\

$^{56}$Cr$(0^+) \rightarrow $$^{56}$Mn($3^+$) & 2nd FU  & 1626.5(6) &   & $ 9.5 \times 10^{2}$ yr&  $ 4.4(60) \times 10^{2}$ yr &gxpf1a \\

$^{56}$Mn$(3^+) \rightarrow $$^{56}$Fe($0^+$) & 2nd FU  & 3695.50(21)  &     & 4.1 yr & 3.9(6) yr & gxpf1a \\

$^{58}$Co$^*$$(5^+) \rightarrow $$^{58}$Ni($0^+$) & 4th FU  & 381.6(11)  &  &  2.0$\times 10^{18}$ yr & 1.3(2)$\times 10^{18}$ yr &gxpf1a \\

$^{60}$Fe$(0^+) \rightarrow ^{60}$Co($5^+$) & 4th FU  & 237(3)  &    &2.2$\times 10^{19}$ yr& 2.5(4)$\times 10^{19}$ yr& gxpf1a  \\

$^{60}$Co$(5^+) \rightarrow ^{60}$Ni($0^+$) & 4th FU  & 2822.81(21) & &  $2.4\times 10^{9}$ yr& $3.1(5)\times 10^{9}$ yr&gxpf1a \\

$^{62}$Fe$(0^+) \rightarrow ^{62}$Co$^*$($5^+$) & 4th FU  & 2546(19) &   &$2.9\times 10^{9}$ yr  &   $1.9(4)\times 10^{9}$ yr & gxpf1a \\

$^{62}$Co$^*$$(5^+) \rightarrow $$^{62}$Ni($0^+$) & 4th FU  & 5322 (19)&   &  $1.1\times 10^{7}$ yr  & $4.7(6)\times 10^{12}$ yr&gxpf1a \\

\hline
$^{84}$Se$(0^+) \rightarrow ^{84}$Br$^*$($6^-$) & 5th FU  &1835(26)  &    & $8.5\times 10^{17}$ yr & $2.7(8)\times 10^{19}$ yr
& gwbxg \\
$^{84}$Br$^*$$(6^-) \rightarrow $$^{84}$Kr($0^+$) & 5th FU  & 4656(26)  &    & $9.0\times 10^{9} $ yr &$1.7(3)\times 10^{10} $ yr  & gwbxg \\

$^{84}$Rb$^*$$(6^-) \rightarrow $$^{84}$Sr($0^+$) & 5th FU  & 890.6(23)&   & $1.1\times 10^{21} $ yr & $1.7(3))\times 10^{17} $ yr & gwbxg \\

$^{86}$Rb$^*$$(6^-) \rightarrow $$^{86}$Sr($0^+$) & 5th FU  &1776.10(20)  &     &$4.3\times 10^{14} $ yr &$3.3(5)\times 10^{14} $ yr &gwbxg \\

$^{92}$Nb$(7^+) \rightarrow ^{92}$Mo($0^+$) & 6th FU  & 355.3(18)   &   & $5.8\times 10^{28} $ yr &  $6.3(1)\times 10^{28} $ yr&  gwbxg \\

$^{94}$Nb$^*$$(3^+) \rightarrow $$^{94}$Mo($0^+$) & 2nd FU  & 2045.0(15)&     & $1.1 \times 10^{2}$yr & $1.7(30) \times 10^{2} $yr & gwbxg \\

$^{96}$Tc$(7^+) \rightarrow ^{96}$Ru($0^+$) & 6th FU  & 259(5) &   &$4.3\times 10^{30} $ yr  & $1.7(4)\times 10^{30} $ yr &  gwbxg  \\

$^{98}$Nb$^*$$(5^+) \rightarrow $$^{98}$Mo($0^+$) & 4th FU  &4591(5) &     & $3.0\times 10^{7} $ yr  &$3.0(5)\times 10^{6} $ yr  & gwbxg \\

$^{100}$Nb $^*$$(5^+) \rightarrow $$^{100}$Mo($0^+$) & 4th FU  &6402(8) &    & $4.3\times 10^{4} $  yr&  $6.9(9)\times 10^{4} $  yr& gwbxg \\

$^{104}$Mo$(0^+) \rightarrow ^{104}$Tc($3^+$) & 2nd FU  & 2155(24)  &  &   $1.6\times 10^{2} $  yr &  $4.1(6)\times 10^{1} $yr & gwbxg \\

$^{104}$Tc$(3^+) \rightarrow ^{104}$Ru($0^+$) & 2nd FU  & 5597(25)&      &  2.3 yr & 78.1(10) d & gwbxg   \\

$^{104}$Rh$^*$$(5^+) \rightarrow $$^{104}$Pd($0^+$) & 4th FU  & 2435.8(27)   &      & $4.4 \times 10^{11} $ yr   & $2.5(4)\times 10^{9} $ yr & gwbxg  \\

\hline
$^{98}$Zr$(0^+) \rightarrow ^{98}$Nb$^*$($5^+$) & 4th FU  & 2243(10)&    & $8.3\times 10^{8} $  yr & $1.7(3)\times 10^{9} $  yr&G-matrix  \\

$^{100}$Zr$(0^+) \rightarrow $$^{100}$Nb$^*$($5^+$) & 4th FU  &3419(11) &   & $9.1\times 10^{7} $  yr   & $3.1 (6)\times 10^{7} $  yr& G-matrix \\
				
\end{tabular}
\end{ruledtabular}
\end{table*}
	%
\begin{table*}
\caption{ Same as Table \ref{table1}.}

\begin{ruledtabular}
			
\begin{tabular}{lccccccc}
\label{table2}

~~~~~Transition &Forbiddenness & Q value & \multicolumn{3}{c}{Partial half-life}   &Interaction \\

\hline
			
~~~~~$\beta^-$ Decay & Kth FU &(keV)  &  &SM& Ref. \cite{joel12017} & \\

\hline

$^{118}$Cd$(0^+) \rightarrow $$^{118}$In$^*$($5^+$) & 4th FU  & 527(21)  &   & $2.5 \times 10^{16} $ yr &  $1.8(5) \times 10^{16} $ yr& G-matrix \\

$^{118}$Cd$(0^+) \rightarrow $$^{118}$In$^*$($8^-$) & 7th FU  & 527(21)  &  & $8.6 \times 10^{33} $ yr & $3.1(1 $mag$) \times 10^{32} $ yr & G-matrix \\

$^{120}$Cd$(0^+) \rightarrow $$^{120}$In$^*$($5^+$) & 4th FU  &1770(40)  &  & $7.0 \times 10^{10} $ yr  & $3.8(8)\times 10^{10} $ yr & G-matrix \\

$^{120}$Cd$(0^+) \rightarrow $$^{120}$In$^*$($8^-$) & 7th FU  & 1770(40) &    &$5.5 \times 10^{24} $ yr  & $2.2(3) \times 10^{22} $ yr & G-matrix \\

$^{120}$Pd$(0^+) \rightarrow $$^{120}$Ag$^*$($6^-$) & 5th FU  & 5372(5)  &     & $1.4\times 10^{10} $ yr& $4.4(7) \times 10^{8} $ yr  &G-matrix \\

$^{120}$Pd$(0^+) \rightarrow $$^{120}$Ag($3^+$) & 2nd FU  & 5372(5)  &  &   $30.4 $  d   & $8.8 (6)$ d & G-matrix \\

$^{120}$Ag$(3^+) \rightarrow $$^{120}$Cd($0^+$) & 2nd FU  & 8306 (6)&  &  $6.4$ d  &  $4.5(8)$ d &G-matrix \\

$^{120}$Ag$^*$$(6^-) \rightarrow $$^{120}$Cd($0^+$) & 5th FU  & 8306(6)  &  &  $9.0 \times 10^{7} $ yr  & $2.2(4) \times 10^{6} $ yr &G-matrix \\

$^{122}$Cd$(0^+) \rightarrow $$^{122}$In$^*$($8^-$) & 7th FU  &2960(50)&   & $8.6\times 10^{19} $ yr   & $2.0(7)\times 10^{19} $ yr  &  G-matrix \\
$^{122}$Cd$(0^+) \rightarrow $$^{122}$In$^*$($5^+$) & 4th FU  &2960(50) &     & $3.2\times 10^{7} $ yr  &  $2.0(4)\times 10^{8} $ yr & G-matrix \\

$^{124}$Cd$(0^+) \rightarrow $$^{124}$In$^*$($8^-$) & 7th FU  & 4170(30) &    & $1.8\times 10^{17} $ yr & $9.4(2)\times 10^{15} $ yr &G-matrix  \\

\hline
$^{114}$In$^*$$(5^+) \rightarrow $$^{114}$Sn($0^+$) & 4th FU  &1989.9(3) &   &$2.6 \times 10^{12} $ yr   & $9.4(2) \times 10^{9} $ yr & snet \\

$^{116}$In$^*$$(5^+) \rightarrow $$^{116}$Sn($0^+$) & 4th FU  & 3276.22(24)&   & $3.8\times 10^{9} $ yr  & $1.7(3) \times 10^{13} $ yr   & snet \\

$^{116}$In$^*$$(8^-) \rightarrow $$^{116}$Sn($0^+$) & 7th FU  &  3276.22(24) &    & $5.4 \times 10^{20} $ yr   & $1.4(2) \times 10^{30} $ yr &snet \\

$^{118}$In$^*$$(5^+) \rightarrow $$^{118}$Sn($0^+$) & 4th FU  & 4425(8)&      & $1.3 \times 10^{8} $ yr & $1.7(3) \times 10^{12} $ yr & snet \\

$^{118}$In$^*$$(8^-) \rightarrow $$^{118}$Sn($0^+$) & 7th FU  & 4425(8) &   &  $3.1 \times 10^{18} $ yr & $4.2(5) \times 10^{20} $ yr &snet \\

$^{120}$In$^*$$(5^+) \rightarrow $$^{120}$Sn($0^+$) & 4th FU  & 5370(40)&     & $1.2 \times 10^{7} $ yr & $4.0(6) \times 10^{17} $ yr & snet \\


$^{122}$In$^*$$(5^+) \rightarrow $$^{122}$Sn($0^+$) & 4th FU  & 6370(50)&   & $2.1\times 10^{6} $ yr  & $1.1(2)\times 10^{14} $ yr &snet \\

$^{122}$In$^*$$(8^-) \rightarrow $$^{122}$Sn($0^+$) & 7th FU  &6370(50) &  & $4.1 \times 10^{15} $ yr & $4.7(2) \times 10^{18} $ yr &snet \\

$^{124}$In$^*$$(8^-) \rightarrow $$^{124}$Sn($0^+$) & 7th FU  & 7360(30)&     &$6.4 \times 10^{14} $ yr & $5.3(8) \times 10^{18} $ yr &snet \\

\hline
$^{126}$Sn$(0^+) \rightarrow $$^{126}$Sb$^*$($5^+$) & 4th FU  & 378(30)&     & $2.6 \times 10^{18} $ yr  & $4.7$(1 mag) $\times 10^{18} $ yr & sn100pn \\

$^{126}$Sn$(0^+) \rightarrow $$^{126}$Sb($8^-$) & 7th FU  &378(30) &   &$3.7 \times 10^{33} $ yr & $2.3(4) \times 10^{15} $ yr &sn100pn \\

$^{128}$Sn$(0^+) \rightarrow $$^{128}$Sb$^*$($5^+$) & 4th FU  & 1268(13)&   & $1.2\times 10^{13} $ yr  &  $2.8(4)\times 10^{13} $ yr & sn100pn \\

$^{128}$Sn$(0^+) \rightarrow $$^{128}$Sb($8^-$) & 7th FU  &1268(13) & & $1.1\times 10^{26} $ yr  &  $1.4(3)\times 10^{23} $ yr &sn100pn \\

$^{130}$Sn$(0^+) \rightarrow $$^{130}$Sb($8^-$) & 7th FU  &2153(14) &   &  $2.3\times 10^{22} $ yr  &  $3.1(5)\times 10^{19} $ yr & sn100pn \\

$^{120}$Sb$^*$$(8^-) \rightarrow $$^{120}$Te($0^+$) & 7th FU  & 945(7) &   & $8.3\times 10^{29} $ yr  &  $4.7(1)\times 10^{26} $ yr& sn100pn \\

$^{122}$Sb$^*$$(8^-) \rightarrow $$^{122}$Te($0^+$) & 7th FU  & 1979.1(21)&    & $2.8\times 10^{24} $ yr & $4.7(6)\times 10^{21} $ yr & sn100pn \\

$^{124}$Sb$^*$$(5^+) \rightarrow $$^{124}$Te($0^+$) & 4th FU  & 2905.07(13)&      & $9.0\times 10^{10} $ yr & $1.9(3)\times 10^{12} $ yr & sn100pn \\

$^{124}$Sb$^*$$(8^-) \rightarrow $$^{124}$Te($0^+$) & 7th FU  &2905.07(13) &    & $1.8\times 10^{22} $ yr& $3.4(5)\times 10^{19} $ yr & sn100pn \\

$^{126}$Sb$^*$$(5^+) \rightarrow $$^{126}$Te($0^+$) & 4th FU  &3670(30)  &  & $6.7\times 10^{9} $ yr   & $1.1(2)\times 10^{11} $ yr & sn100pn \\

\end{tabular}
\end{ruledtabular}
\end{table*}

\begin{table*}
\caption{Same as Table \ref{table1}.}
		%
\begin{ruledtabular}
	\label{table3}		
\begin{tabular}{lccccccccccc}

~~~~~Transition &Forbiddenness & Q value & \multicolumn{3}{c}{Partial half-life}   &Interaction \\

\hline
			
~~~~~$\beta^-$ Decay & Kth FU &(keV)  &   &SM& Ref.\cite{joel12017} & \\

\hline

$^{126}$Sb$(8^-) \rightarrow $$^{126}$Te($0^+$) & 7th FU  &3670(30) &  & $5.0\times 10^{20} $ yr & $1.1(2)\times 10^{18} $ yr & sn100pn \\

$^{128}$Sb$^*$$(5^+) \rightarrow $$^{128}$Te($0^+$) & 4th FU  &4364(19) &    & $1.0\times 10^{9} $ yr &  $1.6(2)\times 10^{10} $ yr& sn100pn \\

$^{130}$Sb$(8^-) \rightarrow $$^{130}$Te($0^+$) & 7th FU  &5067(14) & & $4.0\times 10^{18} $ yr & $9.4(2)\times 10^{15} $ yr & sn100pn \\
   
   $^{132}$Sb$(8^-) \rightarrow $$^{132}$Te($0^+$) & 7th FU  & 5553(4) & & $1.2\times 10^{18} $ yr  &  $4.4(6)\times 10^{15} $ yr& sn100pn \\

$^{134}$Te$(0^+) \rightarrow $$^{134}$I$^*$($8^-$) & 7th FU  &1510(5) &    &$7.6\times 10^{26} $ yr   & $3.1(5)\times 10^{23} $ yr  & sn100pn \\

$^{130}$I$(5^+) \rightarrow $$^{130}$Xe($0^+$) & 4th FU  &2944(3) &   & $1.5\times 10^{11} $ yr & $7.8(1)\times 10^{11} $ yr&sn100pn \\

$^{134}$I$^*$$(8^-) \rightarrow $$^{134}$Xe($0^+$) & 7th FU  &4082(5) &    &  $1.7\times 10^{20} $ yr& $1.1(2)\times 10^{17} $ yr &sn100pn \\

$^{134}$Cs$^*$$(8^-) \rightarrow $$^{134}$Ba($0^+$) & 7th FU  &2058.84(25)&   &  $5.9\times 10^{24} $ yr &  $3.9(5)\times 10^{21} $ yr&sn100pn  \\

$^{136}$Cs$(5^+) \rightarrow $$^{136}$Ba($0^+$) & 4th FU  & 2548.2(19)&   &   $4.8\times 10^{11} $ yr  & $7.8(1)\times 10^{11} $ yr& sn100pn \\

$^{136}$Cs$^*$$(8^-) \rightarrow $$^{136}$Ba($0^+$) & 7th FU  & 2548.2(19)  &   & $5.0\times 10^{23} $ yr  &  $3.1(5)\times 10^{19} $ yr& sn100pn \\

$^{138}$La$(5^+) \rightarrow $$^{138}$Ce($0^+$) & 4th FU  & 1052.42(41) &   &  $6.2\times 10^{15} $ yr & $1.4(3)\times 10^{16} $ yr &sn100pn \\

\hline
$^{136}$Te$(0^+) \rightarrow $$^{136}$I$^*$($6^-$) & 5th FU  & 5120(14) &    &  $2.0\times 10^{8} $ yr & $2.3(5)\times 10^{9} $ yr & jj56pnb \\

$^{138}$Xe $(0^+) \rightarrow $$^{138}$Cs$^*$($6^-$) & 5th FU  & 2915(10) &   &  $3.0\times 10^{12} $ yr& $1.3(3)\times 10^{11} $ yr & jj56pnb  \\

\end{tabular}
\end{ruledtabular}
\end{table*}

\begin{table*}
		\centering
		\label{table4}
		\caption{\color{black}Comparison of shell model half-lives for $\beta^{-}$ transitions  with the  coupling constants $g^{eff}_{A}$=1.0, $g_{V}$=1.0,  and the measured half-lives (or lower limits). The ground-state to ground-state Q value is taken from AME2020 \cite{AME2020}.} 
		\begin{ruledtabular}
			\begin{tabular}{lccccccc}
~~~~~Transition &Forbiddenness & Q value & \multicolumn{2}{c}{Partial half-life (yr)} &Interaction \\

\hline
			
~~~~~$\beta^-$ Decay & Kth FU &(keV)  &  Experimental   &SM & \\

\hline

				
$^{40}$K$(4^-) \rightarrow $$^{40}$Ca($0^+$) & 3rd FU & 1310.91(6) & $1.398(3)\times 10^{9}$  \cite{Belli_2019,Matthew2023}  & $9.2\times 10^{8}$    &sdpf-mu \\

$^{48}$Ca$(0^+) \rightarrow $$^{48}$Sc($5^+$) & 4th FU  & $279(5)$& $>2.5\times 10^{20}$  \cite{Bakalyarov2002}   & $5.1 \times 10^{20}$      &gxpf1a \\

 $^{96}$Zr$(0^+) \rightarrow ^{96}$Nb($5^+$) & 4th FU  & 163.97(10) &$>2.4\times 10^{19} $   \cite{96Zr}  & $1.3\times 10^{20} $   &gwbxg \\

 $^{138}$La$(5^+) \rightarrow $$^{138}$Ce($2^+$) & 2nd FU  &  1052.5(4)& $2.97(3) \times 10^{11}$ \cite{Belli_2019,Matthew2023}  &$7.7\times 10^{10} $   &sn100pn \\
\end{tabular}
\end{ruledtabular}
\end{table*}

 \section{Results and Discussion} \label{result}

{\color{black}
In the present manuscript, we have calculated the SM half-lives for higher forbidden unique $\beta^{-}$ decay transitions. The ground state to ground state  Q values are adopted from  AME2020 \cite{AME2020}. 
The SM half-lives are calculated for $g_A^{eff}=1.0$ and $g_V=1.0$ and  compared with the scaled half-lives from pnQRPA approach for $g_A^{eff}=1.0$ which are extracted from Ref. \cite{joel12017}, see Tables \ref{table1}-\ref{table3}. \\
For mass range $A=50-62$, the OBTDs are calculated using gxpf1a interaction without truncation. Table \ref{table1} clearly shows that as forbiddenness increases, the half-lives also increase by several orders of magnitude. 
The half-lives vary from an hour to several years. For example, the SM half-life for $^{52}$Sc$(3^+)\rightarrow $$^{52}$Ti($0^+$) is 54.3 h and for $^{60}$Fe$(0^+) \rightarrow $$^{60}$Co($5^+$) is $2.2\times 10^{19}$ year, respectively and these half-lives are also comparable with  scaled half-lives from pnQRPA approach.
In the case of $^{58}$Co*, the involved excited isomeric state is represented by an asterisk$(*)$, similar notation has been used for other isomeric states. 
For $^{52}$V and $^{62}$Co*, we see a large difference in the order of half-lives from SM and pnQRPA approach. 
For $^{58}$Co$^*$$(5^+) \rightarrow $$^{58}$Ni($0^+$) and   $^{60}$Fe$(0^+) \rightarrow $$^{60}$Co($5^+$) transitions, the  half-lives are longer as compared to other 4th- FU transitions in the mass range A=50-62 because of small Q-values involved.}

{\color{black}We have employed gwbxg effective interaction for the $^{84}$Se to $^{104}$Rh$^*$.  From $^{84}$Se to $^{92}$Nb, in our SM calculations, we filled $2p_{1/2}$ neutron orbital to make $^{68}$Ni as a core and there is further one neutron excitation from 1$g_{9/2}$ to  1$g_{7/2}$, 2$d_{5/2}$, 2$d_{3/2}$, and 3$s_{1/2}$ orbitals. From $^{94}$Nb$^*$ to  $^{104}$Rh$^*$, we have completely filled the lower orbitals 2$p_{1/2}$, 1$g_{9/2}$ to make $^{78}$Ni as a core. 
For the cases of $^{84}$Se, $^{84}$Rb$^*$, $^{104}$Tc and $^{104}$Rh$^*$, we obtain a large difference in the half-lives between SM and pnQRPA approaches.

 We have performed SM calculations without any truncations, using the G-matrix effective interaction for $^{98}$Zr to $^{124}$Cd. For most of the transitions, we obtain larger SM half-lives compared to the scaled  half-lives from the pnQRPA. However, in the cases of  $^{98}$Zr$(0^+)$$ \rightarrow $$^{98}$Nb$^*$($5^+$) and 
 $^{122}$Cd$(0^+)$$ \rightarrow $$^{122}$In$^*$($5^+$) transitions, the SM half-lives are shorter than the scaled pnQRPA half-lives.}
 
{\color{black}In the Indium chain, we have performed the SM calculations using snet effective interaction. Due to large dimensions, we have restricted the neutron orbitals. For $^{114}$In and $^{116}$In we have applied a truncation of $\nu h^{(0-2)}_{11/2}$, for $^{118}$In and $^{120}$In, a truncation of  $\nu h^{(0-4)}_{11/2}$ and for $^{122}$In and $^{124}$In, a truncation of  $\nu h^{(0-6)}_{11/2}$, respectively.} Here, all the decay transitions are from the isomeric excited states of odd-odd nuclei to g.s. of an even-even  nuclei.  {\color{black} The SM half-lives for all In isotopes are shorter than the scaled pnQRPA half-lives except for $^{114}$In$^*$$(5^+) \rightarrow $$^{114}$Sn($0^+$) transition.}

We have performed the {\color{black} SM } calculations using sn100pn effective interaction for the  decay transitions in the cases of $^{126-130}$Sn, $^{120-132}$Sb, $^{134}$Te, $^{130,134}$I, $^{134,136}$Cs, and $^{138}$La nuclei. We have not employed any truncation except for $^{120,122}$Sb. {\color{black} A minimum of 6-8 neutrons in the  1$g_{7/2}$ orbital ($\nu g^{(6-8)}_{7/2}$) are restricted in the cases for $^{120,122}$Sb.  
The SM half-lives are longer in most cases, with a few in reasonable agreement with the scaled pnQRPA half-lives.}

For $^{136}$Te and $^{138}$Xe, we have employed the SM calculations using jj56pnb effective interaction without any truncation. {\color{black} For these nuclei, we are getting a difference of one order of magnitude in their half-lives between SM and pnQRPA approach.}

In Table \ref{table4}, we compare SM half-lives with the measured  half-lives or lower limits for the transitions $^{40}$K$(4^-)\rightarrow^{40}$Ca($0^+$), $^{48}$Ca$(0^+)\rightarrow$$^{48}$Sc($5^+$), $^{96}$Zr$(0^+)\rightarrow^{96}$Nb($5^+$) and $^{138}$La$(5^+)\rightarrow$$^{138}$Ce($2^+$). {\color{black}A detailed discussion and comparison of the SM half-lives with the experimental half-lives for these transitions is reported below.}

$^{40}$K$(4^-)\rightarrow$$^{40}$Ca($0^+$): The $^{40}$K is a naturally occurring nucleus and can decay through the
$\beta^{-}$, $\beta^{+}$, and EC  channels. The $^{40}$K isotope is important for geochronological applications.
 The partial half-life for  $\beta^{-}$ channel is {\color{black} $1.398(3)\times 10^{9}$ yr} with branching ratio $89.28(13)\%$ for the corresponding transition \cite{Belli_2019}. The recent measurements have been made for the spectral shape of the $\beta$-spectra \cite{Carles2007}.
For this transition, the {\color{black} OBTDs} are obtained from the {\color{black} SM} using the  2p-2h truncation corresponding to excitation from $sd$ to $pf$ shell to get $4^-$ state of  $^{40}$K  and $0^+$ of $^{40}$Ca  using sdpf-mu interaction. {\color{black}
The shell model half-life is $9.2\times 10^{8} $ yr for this transition with the values $g^{eff}_{A}=1.0$ and $g_{V}=1.0$. 
The variation of the shell-model partial half-life with the  different $g^{eff}_{A}$ values is shown in Fig. \ref{40K}.}

 $^{48}$Ca$(0^+)\rightarrow $$^{48}$Sc($5^+$): The doubly magic nucleus $^{48}$Ca is unstable against a single beta decay.  The  $^{48}$Ca (0+) can decays to $4^{+}$, $5^{+}$, $6^{+}$ states of the daughter nucleus $^{48}$Sc which is followed by a large change in angular momentum along with a small Q value. The decay to  $4^{+}$ and $6^{+}$ states {\color{black} are FNU decay and  to $5^+$ state is a FU decay}. Previously, Aunola et al. and  Haaranen et al. estimated  partial {\color{black}half-lives} of $1.1^{+0.8}_{-0.6}\times 10^{21}$  yr  \cite{Aunola} and  $5.2^{+1.7}_{-1.3} \times 10^{20} $ g$^{-2}_{A}$ yr \cite{Haaranen2014}, respectively. For the  $^{48}$Ca case, there is a competition between  single $\beta$- and $\beta$$\beta$- decay channel with the observed half life for $\beta$$\beta$- decay channel is  $6.4\times 10^{19} $ yr \cite{Arnold2016}. {\color{black}In the present calculations, the SM half-life corresponding to a single $\beta^{-}$ decay is $5.0\times 10^{20} $ yr for $g^{eff}_{A}=1.0$ and $g_{V}=1.0$ and the experimental limit for the half-life is greater than $2.5\times 10^{20} $ yr given in Ref. \cite{Bakalyarov2002}.}

 \begin{figure}
\centering
		\includegraphics[scale=0.44]{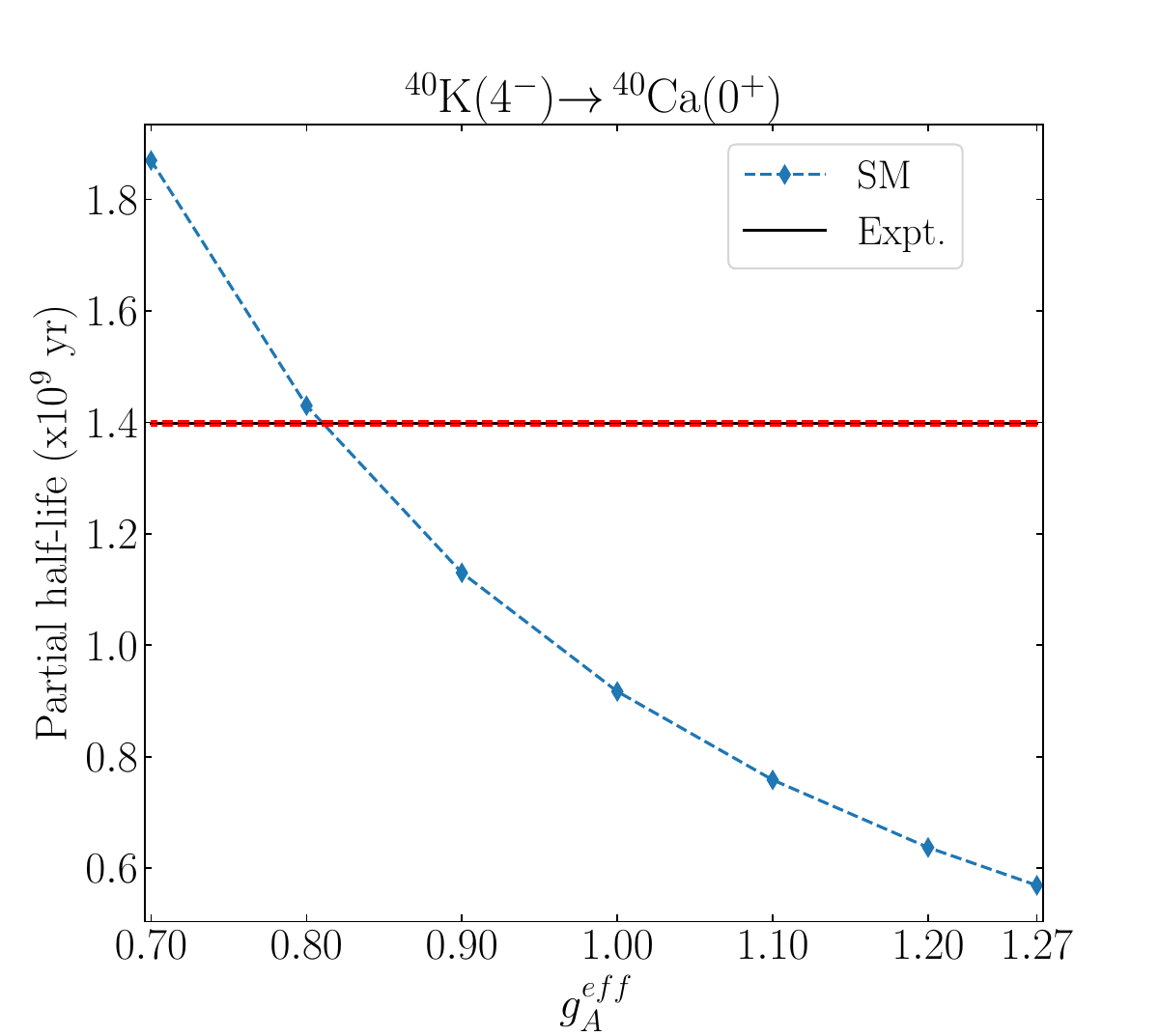}
  \includegraphics[scale=0.44]{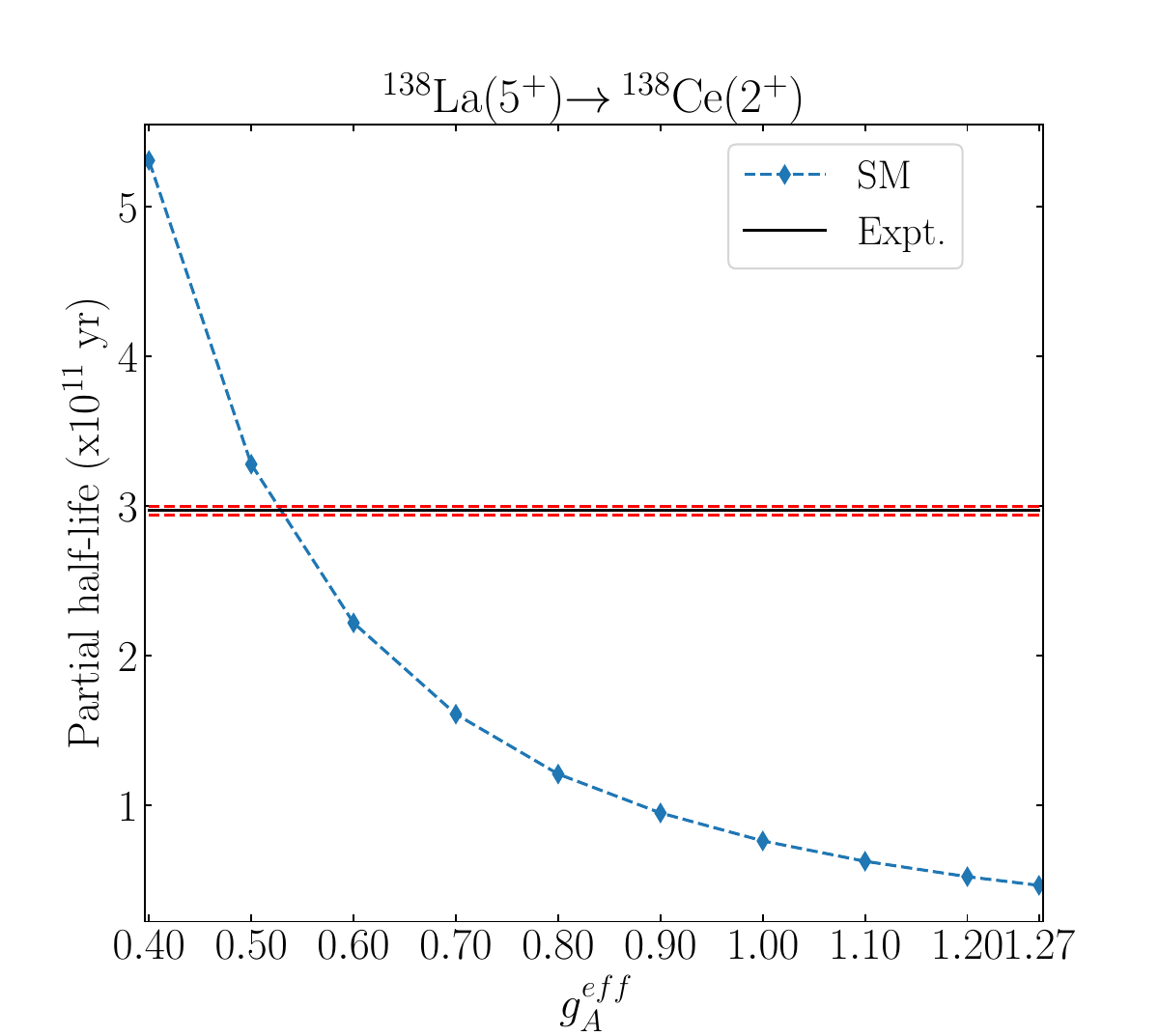}

			\caption{\color{black}The variation of shell model partial half-life with $g_{V}=1.0$ and different $g^{eff}_{A}$ values. The vertical dashed lines (red in colour) are the error value of the experimental partial half-life.\label{40K}}
\end{figure}

$^{96}$Zr$(0^+)\rightarrow $$^{96}$Nb($5^+$): {\color{black}This case is similar to  previous $^{48}$Ca case}. The $0^+$ g.s. can decay to   $4^{+}$,  $5^{+}$ and $6^{+}$ states of daughter nucleus $^{96}$Nb and the most likely branch is {\color{black}4th-} FU decay. The higher forbiddenness 4th- FU decay along with the low Q-value have a longer half-life as compared to the double beta decay channel. The recently measured lower limit of experimental single beta decay half-life is  {\color{black} greater than} $ 2.4 \times 10^{19}$ yr \cite{96Zr}. Previously, calculated shell-model half-life was ($11g^{-2}_{A})\times 10^{19}$ yr \cite {PRL2016} and QRPA half-life was $2.4 \times 10^{20}$ yr \cite{Heiskanen}. {\color{black}In the present work, the calculated shell model half-life with $g^{eff}_{A}=1.0$ and $g_{V}=1.0$  (g.s. to g.s Q value 163.97(0.10) keV) is $ 1.3 \times 10^{20}$ yr.}

$^{138}$La$(5^+)\rightarrow $$^{138}$Ce($2^+$): $^{138}$La is a naturally occurring nucleus with half-life of {\color{black}$1.02(1)\times 10^{11}$ yr }\cite{Audi2017}. The branching ratio for 2nd- FU  $\beta^{-}$ decay $^{138}$La$(5^+) \rightarrow $$^{138}$Ce$^*$($2^+$) is  $34.5\%$ and the  partial half-life for the corresponding transition is $2.97(3)\times 10^{11}$ yr. {\color{black}The SM calculations predict the half-life value  $7.7\times 10^{10}$ yr with $g^{eff}_{A}$= 1.0 and $g_{V}$=1.0. The variation of shell-model results of partial half-life with the  $g^{eff}_{A}$ values is shown in Fig. \ref{40K}}.
}

\section{Conclusion} \label{Conclusion}
In the present work, we have performed the {\color{black}SM }calculations for higher  $\beta^{-}$ decay transitions including 2nd-, 3rd-, 4th-, 5th-, 6th- and 7th- FU,  for nuclei in mass range A=40-138. We have employed the effective interactions sdpf-mu, gxpf1a, gwbxg, G-matrix, snet, sn100pn, and jj56pnb to obtain the OBTDs. Further, the {\color{black} OBTDs} have been used to calculate the partial half-lives within the framework of  $\beta $-decay theory. The present manuscript is focused on those higher FU decay transitions for which experimental data is unavailable except for the 3rd-FU $^{40}$K$(4^-)\rightarrow^{40}$Ca($0^+$), 4th- FU, $^{48}$Ca$(0^+)\rightarrow$$^{48}$Sc($5^+$), 4th- FU $^{96}$Zr$(0^+)\rightarrow ^{96}$Nb($5^+$) and 2nd- FU $^{138}$La$(5^+)\rightarrow $$^{138}$Ce($2^+$) transitions. {\color{black}For these transitions, the measured half-life or its lower limit is available. The calculations are performed with $g^{eff}_{A}$=1.0 and $g_{V}$=1.0 and  results are compared with the expected half-lives from pnQRPA approach at $g^{eff}_{A}$=1.0. The variation in half-lives between the two approaches is due to the use of different nuclear structure formalism involved.
The present SM study will be useful to compare upcoming experimental data for half-lives of these higher forbidden $\beta^{-}$ decay transitions.}

 \section*{Acknowledgement}
	This work is supported by a research grant from SERB (India), CRG/2022/005167. We also acknowledge the National Supercomputing Mission (NSM) for providing computing resources of ‘PARAM Ganga’ at the Indian Institute of Technology Roorkee. We would also like to thank Anil Kumar for the useful discussion.


\end{document}